\documentclass{nature}
\usepackage{amssymb, amsmath}  %% one for symbol, one for math
\usepackage{graphicx}
\usepackage{capt-of}
\usepackage{multirow}
\usepackage{array}
\usepackage[displaymath, mathlines]{lineno}
\linenumbers
\title{Deep Cytometry: Deep learning with Real-time Inference in Cell Sorting and Flow Cytometry}

\author{Yueqin Li$\,^{1,2,3}$, Ata Mahjoubfar$\,^{1,2}$, Claire Lifan Chen$\,^{1,2}$, Kayvan Reza Niazi$\,^{2,4,5}$, Li Pei$\,^3$, \& Bahram Jalali$\,^{1,2,5,6,*}$}  %% '^{}' to make superscripts, '\,' to type commas, '\&' to type &

\let\oldequation\equation
\let\oldendequation\endequation
\renewenvironment{equation}
  {\linenomathNonumbers\oldequation}
  {\oldendequation\endlinenomath}
\let\oldalign\align
\let\oldendalign\endalign
\renewenvironment{align}
  {\linenomathNonumbers\oldalign}
  {\oldendalign\endlinenomath}
\let\oldcenter\center
\let\oldendcenter\endcenter
\renewenvironment{center}
  {\linenomathNonumbers\oldcenter}
  {\oldendcenter\endlinenomath}

\begin{document}
\nolinenumbers  % this removes line numbers, a requirement by arxiv

\maketitle

\begin{affiliations}
\item Department of Electrical \& Computer Engineering, University of California, Los Angeles, California 90095, USA
\item California NanoSystems Institute, Los Angeles, California 90095, USA
\item Key Lab of All Optical Network \& Advanced Telecommunication Network, Ministry of Education, Institute of Lightwave Technology, Beijing Jiaotong University, Beijing 100044, China
\item NantWorks, LLC, Culver City, California 90232, USA
\item Department of Bioengineering, University of California, Los Angeles, California 90095, USA
\item Department of Surgery, UCLA Geffen School of Medicine, Los Angeles, USA

\textsuperscript{*}Correspondence should be addressed to jalali@ucla.edu
\end{affiliations}

\begin{abstract}
Deep learning has achieved spectacular performance in image and speech recognition and synthesis. It outperforms other machine learning algorithms in problems where large amounts of data are available. In the area of measurement technology, instruments based on the photonic time stretch have established record real-time measurement throughput in spectroscopy, optical coherence tomography, and imaging flow cytometry. These extreme-throughput instruments generate approximately 1 Tbit/s of continuous measurement data and have led to the discovery of rare phenomena in nonlinear and complex systems as well as new types of biomedical instruments. Owing to the abundance of data they generate, time-stretch instruments are a natural fit to deep learning classification. Previously we had shown that high-throughput label-free cell classification with high accuracy can be achieved through a combination of time-stretch microscopy, image processing and feature extraction, followed by deep learning for finding cancer cells in the blood. Such a technology holds promise for early detection of primary cancer or metastasis. Here we describe a new deep learning pipeline, which entirely avoids the slow and computationally costly signal processing and feature extraction steps by a convolutional neural network that directly operates on the measured signals. The improvement in computational efficiency enables low-latency inference and makes this pipeline suitable for cell sorting via deep learning. Our neural network takes less than a few milliseconds to classify the cells, fast enough to provide a decision to a cell sorter for real-time separation of individual target cells. We demonstrate the applicability of our new method in the classification of OT-II white blood cells and SW-480 epithelial cancer cells with more than 95\% accuracy in a label-free fashion. 
\end{abstract}

\section*{}
Deep learning provides a powerful set of tools for extracting knowledge that is hidden in large-scale data. In image classification and speech recognition, deep learning algorithms have already made big inroads scientifically and commercially, creating new opportunities in medicine and bioinformatics\cite{min2017deep}. In medicine, deep learning has been used to identify pulmonary pneumonia using chest X-ray images\cite{rajpurkar2017chexnet}, heart arrhythmias using electrocardiogram data\cite{rajpurkar2017cardiologist}, and malignant skin lesions at accuracy levels on par with trained dermatologists\cite{esteva2017dermatologist}. The predictive potential of deep neural networks is also revolutionizing related fields like genetics and biochemistry where the sequence specificities of DNA- and RNA-binding proteins have been determined algorithmically from extremely large and complex datasets\cite{alipanahi2015predicting}. Recently, a deep-learning assisted image-activated sorting technology was demonstrated\cite{nitta2018intelligent}. It used frequency-division-multiplexed microscope to acquire fluorescence image by labeling samples and successfully sorted microalgal cells and blood cells. Moreover, deep learning models helped to analyze water samples so that the ocean microbiome is monitored\cite{gorocs2018deep}.

Flow cytometry is a biomedical diagnostics technique which generates information gathered from the interaction of light (often lasers) with streaming cellular suspensions to classify each cell based on its size, granularity, and fluorescence characteristics through the measurement of forward- and side- scattered signals (elastic scatterings), as well as emission wavelength of fluorescent biomarkers used as marker-specific cellular labels (inelastic scatterings)\cite{shapiro2005practical,watson2004introduction}. One application of this technology is fluorescence-activated cell sorting (FACS) which enables the physical collection of cells of interest away from undesired cells within a heterogeneous mixture using multiple fluorescent labels to apply increasingly stringent light scattering and fluorescent emission characteristics to identify and collect target cell populations. 

Despite the growing utility of flow cytometry in biomedical research and therapeutics manufacturing, the use of this platform can be limited due to the use of labeling reagents which may alter the behavior of bound cells through their inadvertent activation or inhibition prior to collection or through the targeting of unreliable markers for cell identification. CD326/EpCAM\cite{gires2009abundance} is one example of the latter. This protein was initially accepted as a generic biomarker for cancer cells of epithelial origin (or their derivatives such as circulating tumor cells) but was later found to be heterogeneously expressed on both or even absent on the most malignant CTC\cite{kling2012beyond} demonstrating some limitations to this approach. While these findings provide a rationale for the development of label-free cellular analysis and sorting platforms, sole reliance on forward- and side- scattered signals in the absence of fluorescence labeling information has been challenging as a cellular classification modality due to poor sensitivity and selectivity.

As a solution, label-free cell sorting based on additional physical characteristics has gained popularity\cite{shields2015microfluidic,gossett2010label}. This approach is compatible with flow cytometry, but entails rapid data analysis and multiplexed feature extraction to improve classification accuracy. To achieve feature expressivity, parallel quantitative phase imaging (TS-QPI) methods are employed\cite{ikeda2005hilbert,popescu2011quantitative,pham2013real,wei201528} to assess additional parameters such as cell protein concentration (correlated with refractive index) and categorize unlabeled cells with increased accuracy.

We have recently introduced a novel imaging flow cytometer that analyzes cells using their biophysical features\cite{chen2016deep}. Label-free imaging is implemented by quantitative phase imaging\cite{mahjoubfar2013label,chen2014hyper} and the trade-off between sensitivity and speed is mitigated by using amplified time-stretch dispersive Fourier transform\cite{mahjoubfar2017time,goda2013dispersive,mahjoubfar2015design,chen2015optical,solli2009optical,mahjoubfar2013optically,goda2009theory,xing2015ultrafast}. In time-stretch imaging\cite{goda2009serial,goda2012hybrid}, the target cell is illuminated by spatially dispersed broadband pulses, and the spatial features of the target are encoded into the pulse spectrum in a short pulse duration of sub-nanoseconds. Both phase and intensity quantitative images are captured simultaneously, providing abundant features including protein concentration, optical loss, and cellular morphology\cite{feinerman2008variability,sigal2006variability,roggan1999optical,vona2000isolation}. This procedure was successfully used as a classifier for \textit{OT-II} hybridoma T-lymphocytes and \textit{SW-480} colon cancer epithelial cells in mixed cultures and distinct sub-populations of algal cells with immediate ramifications for biofuel production\cite{chen2016deep}. However, the signal processing pipeline to form label-free quantitative phase and intensity images and the image processing pipeline to extract morphological and biophysical features from the images have proven costly in time, taking several seconds to extract the features of each cell\cite{mahjoubfar2017artificial}. This relatively long processing duration prevented the further development of a time-stretch imaging flow cytometer capable of cell sorting because classification decisions need to be made within subseconds, prior to the exit of target cells from the microfluidic channel. Even combined with deep learning methodologies for cell classification following biophysical feature determination, the conversion of waveforms to phase/intensity images and the feature extraction were demanded to generate the input datasets for neural network processing\cite{chen2016deep}.

To remove the time-consuming steps of image formation and hand-crafted feature extraction, we developed and describe the use of a deep convolutional neural network to directly process the one-dimensional time-series waveforms from the imaging flow cytometer and automatically extract the features using the model itself. By eliminating the requirement of an image processing pipeline prior to the classifier, the running time of cell analysis can be reduced significantly. As a result, cell sorting decisions can be made in less than a few milliseconds, orders of magnitude faster than previous efforts\cite{chen2016deep}. Furthermore, we find that some features may not be represented in the phase and intensity images extracted from the waveforms, but can be observed by the neural network when the data is provided as the raw time-series waveforms. These hidden features, not available in manually designed image representations, enhance the model to perform cell classification more accurately. The balanced accuracy and F\textsubscript{1} score of our model reach 95.74\% and 95.71\%, respectively, for an accelerated classifier of \textit{SW-480} and \textit{OT-II} cells, achieving a new state of the art in accuracy, while enabling cell sorting by time-stretch imaging flow cytometry for the first time. Additionally, our technique for real-time processing of signals by deep learning can be used in other optical sensing and measurement systems\cite{jalali2015tailoring,li2017photonic,chen2013ultrafast,yazaki2014ultrafast,li2016theory,mahjoubfar2011high,li2015instantaneous}.

\section*{Results}

In order to better study the learning behavior of the neural network model, the performance of each class and their averaged forms are evaluated for every epoch on the training and validation datasets (Fig. \ref{fig:epochs}). There are multiple ways to measure the performance of the model; tracking the F\textsubscript{1} score is one such example. The F\textsubscript{1} score is the harmonic mean of precision and recall, where precision is the positive predictive value measuring the correctness of the classifier and the recall measures the completeness. Therefore, F\textsubscript{1} score is considered a very effective means of measuring classification performance. Since the examples in the dataset are categorized into three classes (\textit{SW-480}, \textit{OT-II} and blanks), the task for the neural network is multi-class classification as evaluated by calculating the F\textsubscript{1} score per class and also their averaged forms. Three forms of F\textsubscript{1} score averaging are taken into account: (1) the micro-averaged F\textsubscript{1} score, which considers aggregate true positives for precision and recall calculations; (2) the macro-averaged F\textsubscript{1} score, which evaluates precision and recall of each class individually, and then assigns equal weight to each class; (3) and the weighted-averaged F\textsubscript{1} score that assigns a different weight to each class should the dataset be imbalanced. Orange curves show the train F\textsubscript{1} score while green curves show the results of validation F\textsubscript{1} score. Comparing the classification performance for each class, this neural network demonstrates successful recognition of \textit{SW-480} colorectal cells and \textit{OT-II} hybridoma T cells upon completion of the first training epoch. Interestingly, classification of the acellular dataset require approximately 10 epochs to achieve similar performance. The overall performance is determined by the averaged F\textsubscript{1} scores of these three classes. The F\textsubscript{1} scores of the training and validation datasets continue to improve until a maximum is reached at approximately the epoch 60. Meanwhile, the close performance of the train and the validation sets reveals a good generalization of the model. Ultimately, the weighted-averaged validation F\textsubscript{1} score achieved 97.01\%. To evaluate the reproducibility of the results obtained by this neural network, the training procedure was repeated five times starting from randomly initialized weights and biases and demonstrated significant concordance between runs. The standard deviation of the weighted-averaged validation F\textsubscript{1} scores was merely 0.59\% at the last epoch.

% \begin{figure} % To make captionof work, can not use begin{figure} & end{figure} block
% \centering
\begin{center} % Only want the figures and captions in the center, not the other paragraphs.
\includegraphics[scale=0.8]{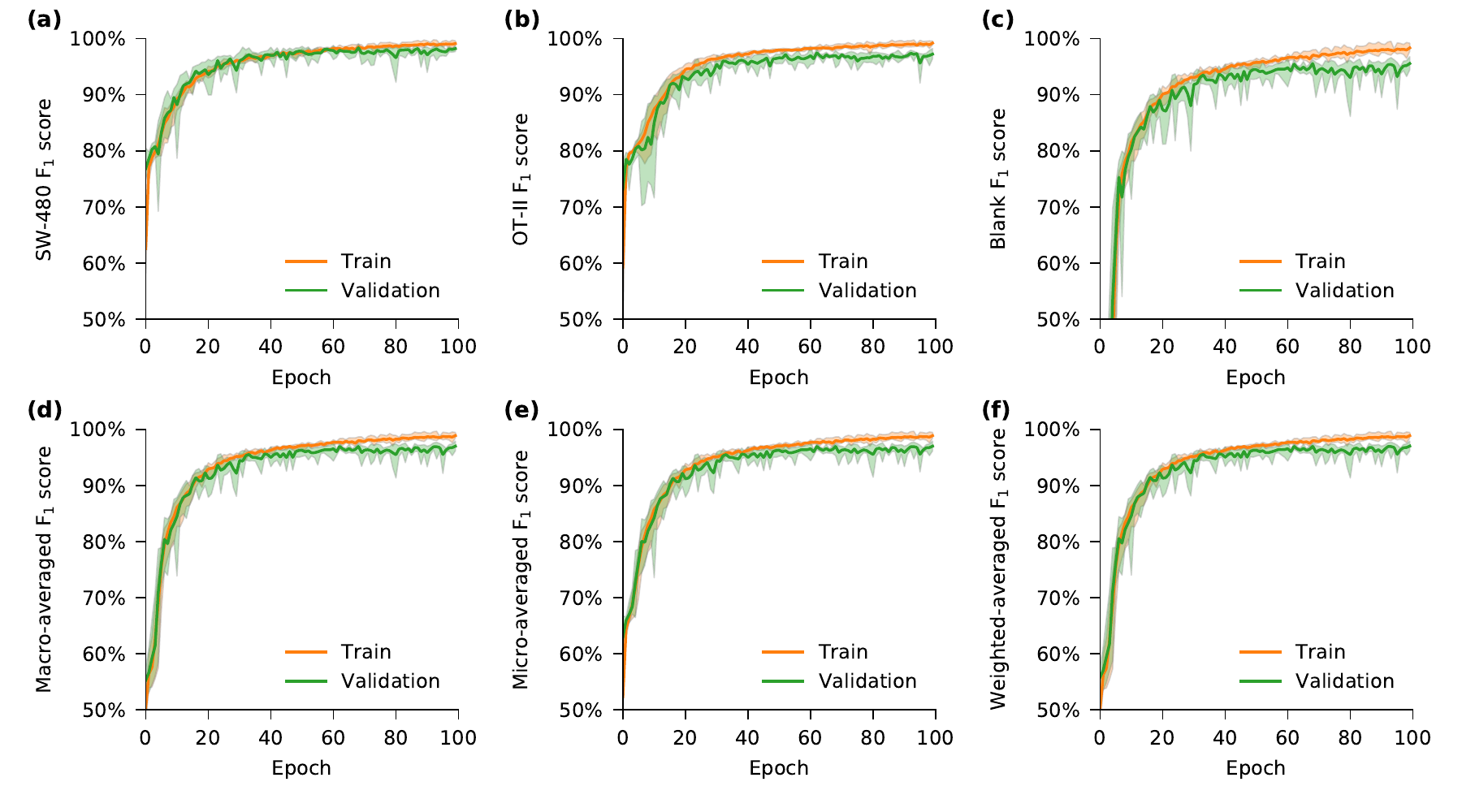} %% When the original figure size is beyond the page size, use scale to adjust it 
\captionof{figure}[caption]{
\label{fig:epochs}
\textbf{Convergence of the network training.}
F\textsubscript{1} score, as a measure of the classification performance, is shown for individual classes (a-c) and their averaged (combined) forms (d-f) over training epochs. At each epoch, the network is trained with all examples in the training dataset, and its performance over these training examples is averaged to obtain the training F\textsubscript{1} score of the epoch (orange curves). At the end of each training epoch, the network is used for classifying all examples in the validation dataset resulting in each epoch's validation F\textsubscript{1} score (green curves). This neural network succeeded to recognize (a) \textit{SW-480} cells and (b) \textit{OT-II} cells even at the end of the first train epoch, but required additional runs to detect (c) regions of the waveform containing no cells (blank examples). The shaded area demonstrates the range of performance variations in each epoch for five different training runs. The validation performance approximates the training performance, indicating the model is well-regularized.\\}
% \end{figure}
\end{center}

\section*{Discussion}

In order for label-free real-time imaging flow cytometry to become a feasible methodology, imaging, signal processing, and data analysis need to be completed while the cell is traveling the distance between the imaging point (field-of-view of the camera) in the microfluidic channel and the cell sorting mechanism (Fig. \ref{fig:sorter}). During imaging, the time-stretch imaging system is used to rapidly capture the spatial information of cells at high throughput. A train of rainbow flashes illuminates the target cells as line scans. The features of the cells are encoded into the spectrum of these optical pulses, representing one-dimensional frames. Pulses are stretched in a dispersive optical fiber, mapping their spectrum to time. They are sequentially captured by a photodetector, and converted to a digital waveform, which can be analyzed by the neural network. The imaging and data capture take less than 0.1 ms for each waveform element, which covers a field-of-view of $25 \mu m$ in the channel direction, often containing only one cell surrounded by the suspension buffer or no cell. So, the delay in making a decision for cell sorting is dominated by the data processing time of the neural network.

\begin{center}
\includegraphics[scale=0.55]{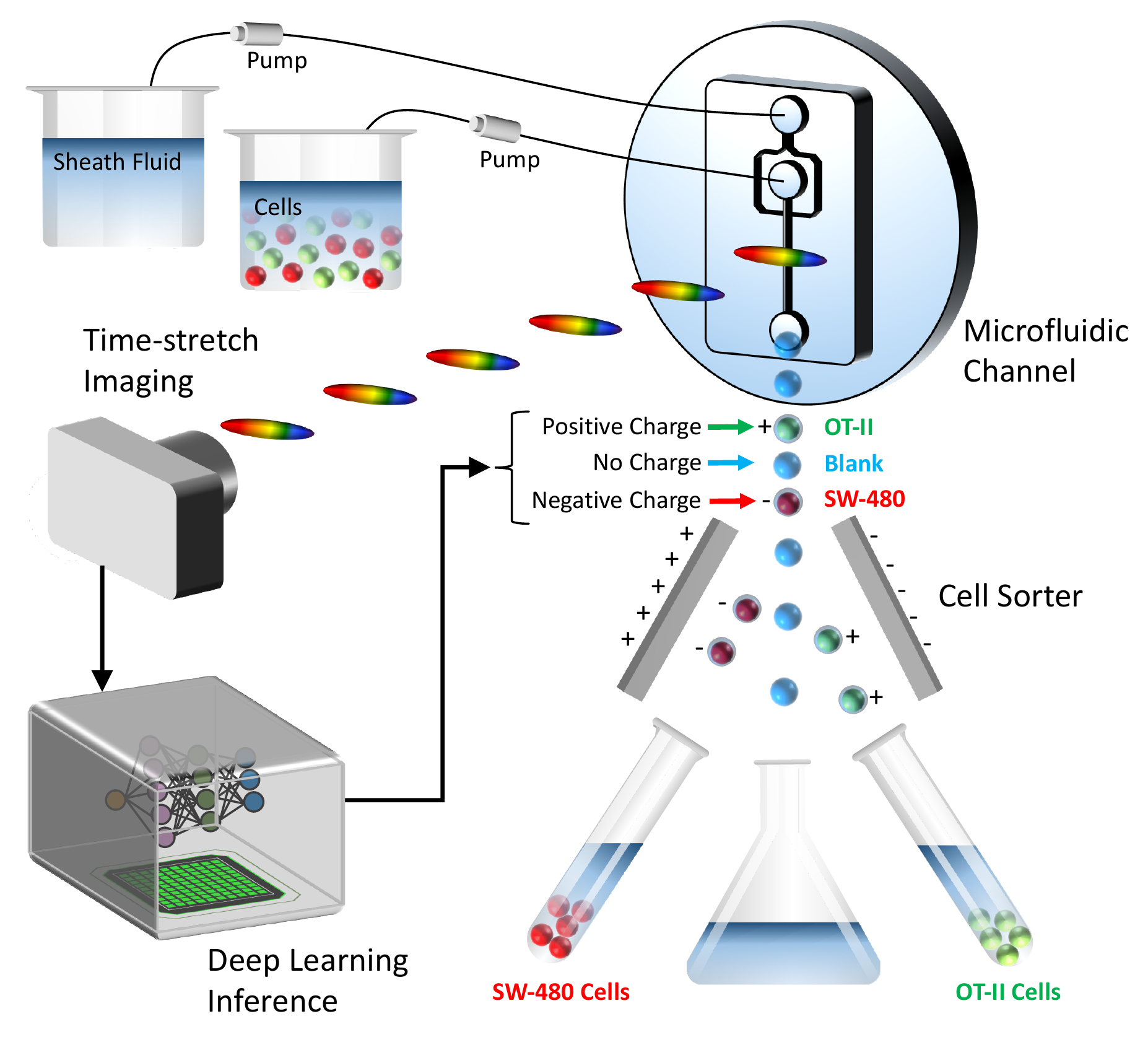} %% when the original figure size is beyond the page size, use scale to adjust it 
\captionof{figure}[caption]{
\label{fig:sorter}
\textbf{Deep cytometry: application of deep learning in cell sorting and flow cytometry.}
A microfluidic channel with hydrodynamic focusing mechanism uses sheath fluid to align the cells in the center of field-of-view. The rainbow pulses formed by the time-stretch imaging system capture line images of the cells in the channel, containing blur-free quantitative label-free images of the cells flowing at a high speed. The output waveforms of the time-stretch imaging system are directly passed to a deep neural network without any signal processing. The network achieves rapid cell classification with high accuracy, fast enough to make decisions before the cells reach the sorting mechanism. Different types of cells are categorized and charged with different polarity charges so that they can be separated into different collection tubes.\\}
% \end{figure}
\end{center}

To quickly classify the target cells based on the collected data, we demonstrate the utility of analyzing waveforms directly by a deep neural network, referred to as deep cytometry. The classification model is trained offline using datasets for the target cell types, and then used in an online system for cell sorting. The processing time of this model (the latency for inference of a single-example batch by a previously trained model) is 23.2 ms per example using an Intel Xeon CPU (8 cores), 8.6 ms per example on an NVIDIA Tesla K80 GPU, and 3.6 ms per example on an NVIDIA Tesla P100 GPU. Thus, for our setup with the cell flow rate of 1.3 m/s in the microfluidic channel, the cells travel 30.2 mm for the Intel CPU, 11.2 mm for the NVIDIA K80 GPU, or 4.7 mm for the NVIDIA P100 GPU before the classification decision is made. So, the microfluidic channels should be at least as long as these cell travel distances. Fabrication of microfluidic channels beyond these length limits is very practical, and the cells can remain ordered within such short distances. Therefore, the type of each cell can be determined by our model in real-time before it reaches the cell sorter. Oftentimes the flow speed is less than our setup, and the length limitation is further relaxed.

\section*{Conclusion}
In this manuscript, a deep convolutional neural network with fast inference for direct processing of flow cytometry waveforms was presented. The results demonstrate record performance in label-free detection of cancerous cells with a test F\textsubscript{1} score of 95.71\% and accuracy of 95.74\% with high consistency and robustness. The system achieves this accurate classification in less than a few milliseconds, opening a new path for real-time label-free cell sorting.

\newpage
\section*{References}
\bibliography{manuscript} 

\begin{addendum}
\item The work was entirely performed at the California NanoSystems Institute at UCLA. This work is partially supported by NantWorks LLC. Y. Li was supported by the China Scholarship Council. B. Jalali would like to thank NVIDIA for the donation of the GPU system.
\item[Author Contributions] Y.L., C.L.C., A.M., and B.J. conceived the idea. Y.L., A.M., and C.L.C. designed the deep convolutional neural network. Y.L., C.L.C., and A.M. performed the experiments, collected the data, and developed the training and evaluation codes. K.R.N. provided the biological cell samples. Y.L., A.M., C.L.C., K.R.N., B.J., and L.P. analyzed the results. Y.L. prepared the figures. Y.L., A.M., B.J., K.R.N., C.L.C., and L.P. wrote and reviewed the manuscript. B.J. supervised the work.
\item[Competing Interests] The authors declare no competing financial and non­financial interests.

\end{addendum}

\end{document}